\documentstyle[12pt,axodraw,epsfig]{article}

\advance\textheight by \topskip
\begin{document}
\tolerance=100000
\thispagestyle{empty}
\setcounter{page}{0}

\newcommand{\be}{\begin{equation}}
\newcommand{\ee}{\end{equation}}
\newcommand{\br}{\begin{eqnarray}}
\newcommand{\er}{\end{eqnarray}}
\newcommand{\ba}{\begin{array}}
\newcommand{\ea}{\end{array}}
\newcommand{\bi}{\begin{itemize}}
\newcommand{\ei}{\end{itemize}}
\newcommand{\bn}{\begin{enumerate}}
\newcommand{\en}{\end{enumerate}}
\newcommand{\bc}{\begin{center}}
\newcommand{\ec}{\end{center}}
\newcommand{\ul}{\underline}
\newcommand{\ol}{\overline}
\def\epem{\ifmmode{e^+ e^-} \else{$e^+ e^-$} \fi}
\newcommand{\eeww}{$e^+e^-\rightarrow W^+ W^-$}
\newcommand{\qqQQ}{$q_1\bar q_2 Q_3\bar Q_4$}
\newcommand{\eeqqQQ}{$e^+e^-\rightarrow q_1\bar q_2 Q_3\bar Q_4$}
\newcommand{\eewwqqqq}{$e^+e^-\rightarrow W^+ W^-\ar q\bar q Q\bar Q$}
\newcommand{\eeqqgg}{$e^+e^-\rightarrow q\bar q gg$}
\newcommand{\eeqqqq}{$e^+e^-\rightarrow q\bar q Q\bar Q$}
\newcommand{\eewwjjjj}{$e^+e^-\rightarrow W^+ W^-\rightarrow 4~{\rm{jet}}$}
\newcommand{\eeqqggjjjj}{$e^+e^-\rightarrow q\bar 
q gg\rightarrow 4~{\rm{jet}}$}
\newcommand{\eeqqqqjjjj}{$e^+e^-\rightarrow q\bar q Q\bar Q\rightarrow
4~{\rm{jet}}$}
\newcommand{\eejjjj}{$e^+e^-\rightarrow 4~{\rm{jet}}$}
\newcommand{\jjjj}{$4~{\rm{jet}}$}
\newcommand{\qqbar}{$q\bar q$}
\newcommand{\ww}{$W^+W^-$}
\newcommand{\ar}{\rightarrow}
\newcommand{\sm}{${\cal {SM}}$}
\newcommand{\Dir}{\kern -6.4pt\Big{/}}
\newcommand{\Dirin}{\kern -10.4pt\Big{/}\kern 4.4pt}
\newcommand{\DDir}{\kern -7.6pt\Big{/}}
\newcommand{\DGir}{\kern -6.0pt\Big{/}}
\newcommand{\wwqqqq}{$W^+ W^-\ar q\bar q Q\bar Q$}
\newcommand{\qqgg}{$q\bar q gg$}
\newcommand{\qqqq}{$q\bar q Q\bar Q$}

\def\Ord{\buildrel{\scriptscriptstyle <}\over{\scriptscriptstyle\sim}}
\def\OOrd{\buildrel{\scriptscriptstyle >}\over{\scriptscriptstyle\sim}}
\def\pl #1 #2 #3 {{\it Phys.~Lett.} {\bf#1} (#2) #3}
\def\np #1 #2 #3 {{\it Nucl.~Phys.} {\bf#1} (#2) #3}
\def\zp #1 #2 #3 {{\it Z.~Phys.} {\bf#1} (#2) #3}
\def\pr #1 #2 #3 {{\it Phys.~Rev.} {\bf#1} (#2) #3}
\def\prep #1 #2 #3 {{\it Phys.~Rep.} {\bf#1} (#2) #3}
\def\prl #1 #2 #3 {{\it Phys.~Rev.~Lett.} {\bf#1} (#2) #3}
\def\mpl #1 #2 #3 {{\it Mod.~Phys.~Lett.} {\bf#1} (#2) #3}
\def\rmp #1 #2 #3 {{\it Rev. Mod. Phys.} {\bf#1} (#2) #3}
\def\sjnp #1 #2 #3 {{\it Sov. J. Nucl. Phys.} {\bf#1} (#2) #3}
\def\xx #1 #2 #3 {{\bf#1}, (#2) #3}
\def\preprint{{\it preprint}}

\begin{flushright}
{\large Cavendish-HEP-95/15}\\ 
{\large DFTT 04/95}\\ 
{\large DTP/95/10}\\ 
{\rm May 1996\hspace*{.5 truecm}}\\ 
\end{flushright}

\vspace*{\fill}

\begin{center}
{\Large \bf Perturbative rates and 
colour rearrangement 
effects in  four--jet events at LEP2
%\footnote{Work supported 
%in part by Ministero dell' Universit\`a e della Ricerca Scientifica.\\[4. mm]
%E-mails:
%Ballestrero@to.infn.it;  V.A.Khoze@durham.ac.uk; Maina@to.infn.it;
%Moretti@hep.phy.cam.ac.uk; W.J.Stirling@durham.ac.uk.}
}\\[1.cm]

{\large A.~Ballestrero$^a$, V.A.~Khoze$^b$, E.~Maina$^a$,}\\[0.25 cm]
{\large S.~Moretti$^{a,c}$ and W.J.~Stirling$^{b,d}$}\\[0.4 cm]
{\it a) Dipartimento di Fisica Teorica, Universit\`a di Torino,}\\
{\it and I.N.F.N., Sezione di Torino,}\\
{\it Via Pietro Giuria 1, 10125 Torino, Italy.}\\
{\it b) Department of Physics, University of Durham,}\\
{\it South Road, Durham DH1 3LE, UK.}\\
{\it c) Cavendish Laboratory, University of Cambridge,}\\
{\it Madingley Road, Cambridge CB3 0HE, UK.}\\
{\it d) Department of Mathematical Sciences, University of Durham,}\\
{\it South Road, Durham DH1 3LE, UK.}\\
\end{center}

\vspace*{\fill}

\begin{abstract}
{\normalsize
\noindent
An important issue in  the direct reconstruction method of
determining the $W$ mass 
from  $q\bar q Q\bar Q$ events at LEP2 concerns the impact of the 
relatively unknown QCD interconnection effects. It has been suggested
that a study of  `short string' states, in 
which colour singlet states
are formed from $q \bar Q $ and $Q\bar q$ pairs with small phase--space
separation, could shed important light on this issue.
We show that  such configurations can also be
generated by  conventional background $e^+e^-\ar 4$~parton processes,
in particular QCD $q \bar q g g $ and $q \bar q Q \bar Q $
and non--resonant electroweak $q \bar q Q \bar Q $ production.
We study the colour and kinematic structure  of these background
contributions, and estimate the event rate to be  expected at LEP2.
We find  that  the QCD processes
are heavily suppressed, but that  non--resonant  $q \bar q Q \bar Q$ 
production may be comparable in rate to the expected `short string' signal
from $W^+W^-$ production.}
\end{abstract}

\vspace*{\fill}
\newpage

\section{Introduction}

The detailed study of  $W^\pm$ boson physics is one of the most
important goals of the LEP2 \epem\ collider.
Among the main objectives is a precision measurement of the mass $M_W$
of the $W$ boson, with the target accuracy $\pm50$~MeV, see for example 
Ref.~\cite{accuracy}.

An obvious requirement for the success of these precise studies is a
high level of reliability of theoretical predictions for the signal
and background contributions to various experimental observables
related to the different methods of measuring $M_{W}$ from the process
\eeww. This, in particular, requires a detailed understanding of the
physical phenomena which
describe the production and decay of $W^\pm$ bosons at LEP2, for
instance, of the effects which arise from the relatively 
large $W^\pm$ boson decay
width  $\Gamma_{W}$ ($\simeq 2.1$~GeV \cite{gamw}). The instability of
the $W^\pm$
bosons can, in principle, strongly modify the standard `stable $W^\pm$'
results. For example, an important role can be played by QCD
radiative interferences (both virtual and real) which interconnect the
production and decay stages, see for example Refs.~\cite{inter1,inter2}.

It is currently believed that the highest precision on the $M_W$
determination is obtained from the method of direct kinematic
reconstruction of $M_{W}$ using the decay channels\footnote{We use the
notation $q$ and $Q$ to distinguish quarks from different $W$ decay.}
\be\label{hdecay}
W^+W^-\ar q_1\bar q_2 Q_3\bar Q_4,
\ee
\be\label{ldecay}
W^+W^-\ar q\bar q\ell\nu_\ell.
\ee
  
However, the direct reconstruction method is not without problems.
For example, to construct the two $W^\pm$'s from the $q_1\bar q_2 Q_3\bar Q_4$
final
state in (\ref{hdecay}) one must, in principle, attribute all observed
hadrons to the `correct' parent $W^\pm$, a procedure which is certainly
affected by relatively unknown QCD interconnection corrections
\cite{inter2}. Since a complete description of these effects is not
possible at present, one has to rely on model predictions rather than 
on exact calculations, for details see 
Refs.~\cite{accuracy,inter2,model1,model2,jekg}.
Detailed experimental studies of  four--jet events in \epem\
annihilation corresponding to the kinematics of process 
(\ref{hdecay}) could provide  important information about
the size of interconnection--related systematic uncertainties in the 
$W$ mass measurements.

There is another challenging reason to study carefully the phenomenon
of QCD interconnection (colour rearrangement) in hadronic $W^+W^-$
events. As  first emphasised in Ref.~\cite{lab}, it could provide a new
laboratory for probing non--perturbative QCD dynamics. The 
fact that different models for the non--perturbative fragmentation
give different predictions \cite{accuracy}
 means that it might be possible to learn
about the structure of the QCD vacuum. For example, one may hope to
distinguish various scenarios by exploiting the difference in the
sensitivity of the reconnection to the event topology
\cite{inter2,model1}. 

Unfortunately to make any progress at all, we have so far had to rely on
models and approximations that are far from perfect. There is a true
limit to our current physics understanding. One unresolved problem
concerns an evident breakdown of the exclusive probabilistic
interpretation of the colour  suppressed interference effects, see 
Refs.~\cite{gamw,inter2}.

Recall that there is an important difference between the perturbative
QCD `radiophysics' picture \cite{pert} and the non--perturbative
fragmentation scenarios like, for example, the Lund string model
\cite{swe}. The latter requires a completely exclusive description: in
the end the \qqQQ\ system must be subdivided into and fragment as two
separate colour singlets, either $q_1\bar q_2$ and $Q_3\bar Q_4$ 
 or $q_1\bar Q_4$ and $Q_3\bar q_2$. The
string model therefore predicts effects that could be observed on
an event--by--event basis. On the other hand, within the perturbative QCD
approach, analogously to  other colour  suppressed interference
effects \cite{pert,fong,long}, the colour rearrangement phenomena can
be viewed only on a completely inclusive basis, when all the
antennae/dipoles are simultaneously active in the particle production.
The fact that the reconnection pieces are not positive--definite
\cite{inter2} reflects their wave interference nature. Therefore, the
recoupling effects should appear on top of a background generated by
standard no--reconnection antennae/colour dipoles. Normally
(for example, for $e^+e^-\ar q\bar q g$) the two pictures give quite
similar overall description; 
differences only become dramatic when dealing with 
small colour  suppressed effects, see Refs.~\cite{pert,fong,long}.
Another open question concerns the interplay between the
perturbative and non--perturbative phases in the reconnection \cite{inter2}.

The issue of the experimental observability of colour reconnection
needs a special detailed consideration. For example, in
Ref.~\cite{inter2} the analysis was concerned mainly with the standard
global event measures where the effects appear to be very small. The change
in the average charged multiplicity is predicted at a level of a
percent or less, and similar conclusions hold for rapidity
distributions, thrust distributions, and so on. This is very likely below
the experimental precision one may expect at LEP2, and so
the effects  may well be unobservable.
There are, however, some other potentially promising approaches, 
e.g. the comparison
of the event properties in fully hadronic and mixed leptonic--hadronic
decays or the comparison with measurements from the $p \bar p$
collider, see for example Refs.~\cite{gamw,inter2}. 
In particular, an interesting vista on the reconnection issue
is provided by Bose--Einstein effects \cite{twe}.

A high--statistics run above the $Z^0Z^0$  threshold would of course
allow an
unambiguous determination of any systematic mass shift, since  the
$Z^0$ mass is already very precisely known from  LEP1. If
the various potential sources of systematic error could be
disentangled, it could also provide a direct observation of reconnection
effects. More generally, $Z^0$ events from LEP1 could be used to predict
a number of properties for $Z^0Z^0$ events, such as the charged
multiplicity distribution. Any sign of deviations would then provide
important information on the reconnection issue.

An interesting proposal which attempts to disentangle the recoupling
phenomenon is discussed in Ref.~\cite{model1}, where it is argued that
 dynamical effects could enhance the reconnection probability for
configurations which correspond to so--called `short strings', i.e.
strings connecting $q_1$ with $\bar Q_4$ (and $\bar q_2$ with  $Q_3$)
when the quark--antiquark pairs are close together in phase space,
equivalently, when the event has a high thrust.
Such configurations would be expected to produce fewer hadrons
than average.  It is estimated that with 10\% probability for recoupling
these reconnected events could be experimentally identified.

However, in order to clearly pin down such a manifestation of colour
rearrangement in process (\ref{hdecay}) in  a realistic LEP2 scenario
further effort is required.
One of the most important questions concerns the size of the non--$W^+W^-$
background.  For example, the
`short string' states discussed above can also be generated
by  conventional $e^+e^-\ar
4$~parton events in specific colour configurations which give rise
to `rapidity gaps', see 
Refs.~\cite{trett,tr,fjo,fme}. Such events provide a natural lower
limit for a recoupling--type signal\footnote{We assume here that
these rapidity gap events would be statistically distinguishable from random
fluctuations in  conventional  $e^+e^-\to$~2~jet production}.

In this study we calculate all the four-parton background
contributions which could give events with large rapidity gaps.
We use the exact lowest--order tree-level matrix elements
to compute the overall four--jet cross sections, and then study
the colour and kinematic structure  of the various processes.
In this way we are able to estimate the  probability of finding
background configurations corresponding to two pairs of partons
in colour singlet states and relatively close together in phase space,
configurations which could give rise to rapidity gaps.

The paper is organised as follows. In the following section 
we list the various signal and background
four--parton processes and discuss their colour properties.
In Section~3 we present numerical studies of the corresponding
cross sections at LEP2 energies. Our conclusions are presented
in Section~4.

\section{Recoupled four--jet events}

As discussed in the Introduction, a natural lower limit for the
colour recoupling signal in  $W^+W^-$ hadronic events is provided by
 $e^+e^-\ar$~4 partons background events with 
sufficiently large rapidity
gaps \cite{trett,tr,fjo,fme}. One can expect that when the
two colour singlet quasi--collinear jet pairs are moving apart in the
centre--of--mass  frame
with large velocities, the production of hadrons is suppressed
in the rapidity  region separating the two colour  singlet systems.
Therefore these rapidity gap events can, in principle, mimic the
`short string' signal of the $W^+W^-$ hadronic events advocated in
Ref.~\cite{model1} (see also \cite{lab}) as a  laboratory
for studying colour recoupling effects.\footnote{Note also 
 that the high thrust selection automatically selects a class
of events in which QCD radiation from the final--state
quarks is suppressed.} 
Note that contrary to the $W^+W^-$
case, where the `short string' signal is suppressed by a small (but
theoretically uncertain) recoupling factor, four partons 
in a rapidity gap configuration are produced almost simultaneously in a 
small space--time interval, and so all the QCD antennae/dipoles are
equally active. In addition, the soft and collinear
singularities inherent in the QCD matrix elements imply that
 selecting four--jet events with a 
 high thrust, which enhances the recoupling signal \cite{model1},
automatically enhances the role of rapidity gap background events.
%\footnote{It certainly brings us into the two--jet region. So
%one has to be very cautious here in separating the rearrangement signal.}.

We would expect \cite{trett,tr} that the global features of rapidity
gap events could be evaluated within the framework of perturbative QCD.
We therefore base our analysis on the 
tree--level matrix elements for the following four--jet processes,
\be\label{proc1} 
e^+e^-\ar W^+W^- \ar q_1\bar q_2 Q_3\bar Q_4,
\ee
\be\label{proc2}  
e^+e^-\ar q_1\bar q_2 g_3g_4,
\ee
\be\label{proc3}  
e^+e^-\ar q_1\bar q_2 Q_3\bar Q_4\qquad
\mbox{(via $g$ propagators)},
\ee
\be\label{proc4}  
e^+e^-\ar q_1\bar q_2 Q_3\bar Q_4\qquad\mbox{(via $\gamma,Z^0$ propagators)},
\ee
described by the Feynman diagrams\footnote{$W^\pm$--propagator
contributions are omitted from the diagrams of Fig.~3a for
process (\ref{proc4}).} shown  in
Figs.~1--3. 
The matrix element for (\ref{proc1}) is calculated using
 the techniques described in \cite{metodo} while
 for (\ref{proc2})--(\ref{proc4})  we use the same {\tt FORTRAN} codes of 
%Refs.~\cite{noiPL,noiNP,noiPr}, 
Refs.~\cite{noiPL,noiNP,noiPr}, 
to which the reader can refer
for details. The results obtained for
process (\ref{proc1}) have been cross-checked with a calculation
using the formalism of Ref.~\cite{ks}.

Configurations corresponding to  two
distinctively separated colour  singlet parton pairs, which should lead
to rapidity gap events,  are illustrated schematically in Fig.~4
for the processes (\ref{proc1})--(\ref{proc4}) listed above:
Fig.~4a for 
process ({\ref{proc2}) (see also Fig.~2), Fig.~4b for process (\ref{proc3})
 (see also Fig.~3a when
the jagged line represents a $g$),
 and Figs.~4c,d for processes (\ref{proc1}) (4d only)
and (\ref{proc4}) (see also Fig.~1, Fig.~3a
when the jagged line represents a $\gamma$ or a $Z^0$, and Fig.~3b).
% Two colour  singlet parton pairs
%could appear in the processes (\ref{proc3}) and (\ref{proc4}) 
%in the leading order in $\frac{1}{N_c^2}$, for
%the process (\ref{proc2}) such a configuration is always colour suppressed.

There is a simple heuristic way to derive the colour 
factors for the production
of a $q \bar q$ pair in a colour singlet ($S$) or  octet ($O$) state. 
In terms of colour matrices, the production
of a singlet state must be proportional to the identity and therefore: 

\begin{center}
\begin{picture}(400,90)
\SetWidth{1.2}

\ArrowLine(80,30)(30,30)
\ArrowLine(30,70)(80,70)
\Line(100,55)(115,55)
\Line(100,45)(115,45)
\GOval(90,50)(30,15)(0){1}
\Text(90,50)[]{ \Large {$S$}}
\Text(140,50)[]{ \Large {=}}
\Text(170,50)[]{ \Large {$A_S$}}
\ArrowLine(250,30)(200,30)
\ArrowLine(200,70)(250,70)
\CArc(250,50)(20,-90,90)

\end{picture}
\end{center}
Here and in the following we will only draw the {\it colour part}
of the full Feynman amplitudes. 
The production of a quark--antiquark pair
in an octet state, on the other hand, must be proportional
to the production of
a gluon state:

\begin{center}
\begin{picture}(400,90)
\SetWidth{1.2}

\ArrowLine(80,30)(30,30)
\ArrowLine(30,70)(80,70)
\Line(100,55)(115,55)
\Line(100,45)(115,45)
\GOval(90,50)(30,15)(0){1}
\Text(90,50)[]{ \Large {$O$}}
\Text(140,50)[]{ \Large {=}}
\Text(170,50)[]{ \Large {$A_O$}}
\ArrowLine(250,30)(200,30)
\ArrowLine(200,70)(250,70)
\CArc(250,50)(20,-90,90)
\Gluon(270,50)(310,50){5}{4}

\end{picture}
\end{center}
These equalities are exact, with $A_S= 1/\sqrt{3}$
and  $A_O= \sqrt{2}$, for all quark and gluon indices if, in the
$RGB$ colour  basis, we use the usual
Gell--Mann matrices and the following set of gluon states:
$ g_1=(R \overline{G} + G \overline{R})/\sqrt{2}$, 
$ g_2= i (R \overline{G} - G \overline{R})/\sqrt{2}$, 
$ g_3= (R \overline{R} - G \overline{G})/\sqrt{2}$, 
$ g_4=(B \overline{R} + R \overline{B})/\sqrt{2}$, 
$ g_5= i (B \overline{R} - R \overline{B})/\sqrt{2}$,
$ g_6=(G \overline{B} + B \overline{G})/\sqrt{2}$, 
$ g_7=i (G \overline{B} - B \overline{G})/\sqrt{2}$, 
$ g_8= (R \overline{R} + G \overline{G} - 2 B \overline{B})/\sqrt{6}$,
with the singlet state  given by
$ g_0= (R \overline{R} + G \overline{G} + B \overline{B})/\sqrt{3}$.
When computing colour traces only $A_S^2$ and $A_O^2$ are needed.
These can also be obtained
from the processes $e^+e^- \rightarrow q \bar q$, which can only result
in a singlet state, and $q_1 \bar q_1\rightarrow q_2 \bar q_2$,
with $q_1\neq q_2$, in which the initial and final
quark pairs are necessarily in the octet state. One again finds
$A_S^2 = 1/3$ and $A_O^2 = 2$. 

An alternative way to obtain the same results is to recall the
identity,
see for example Ref.~\cite{pert},
\begin{equation}
T^a_{ij}T^a_{kl}= \frac{1}{2}\left( \delta_{il}\delta_{jk}-\frac{1}{3}
 \ \delta_{ij}\delta_{kl}\right)
\end{equation}
which graphically reads: 

\begin{center}
\begin{picture}(450,90)
\SetScale{.9}
\SetWidth{1.2}
\SetOffset(50,0)

\ArrowArcn(0,50)(30,90,270)
\Gluon(30,50)(70,50){5}{4}
\ArrowArcn(100,50)(30,270,90)

\Text(110,50)[]{ \Large{= $\frac{1}{2}$ }}

\ArrowLine(145,75)(195,75)
\ArrowLine(195,25)(145,25)

\Text(200,50)[]{ \Large{- $\frac{1}{6}$}}

\ArrowArcn(240,50)(30,90,270)
\ArrowArcn(320,50)(30,270,90)

\end{picture}
\end{center}
By multiplying by two and
isolating the first term on the right--hand side, which represents
a generic $q\bar q$ state, it is straightforward to
show that that the coefficients of the singlet 
 and  octet terms
are exactly $A_S^2$ and $A_O^2$ respectively. 

Using these simple calculational tools one can compute the colour factors
for the production of $q \bar q$ pairs in the singlet state
for all the above four--jet processes in $e^+e^-$ collisions.
It is sufficient to join the quark and antiquark 
{\it colour lines} in the amplitude and then to compute the
colour factor for the amplitude squared with the usual
rules, dividing the final result by three. Each of the 
processes under consideration is discussed in turn below.
\begin{itemize}

\item
$e^+e^-\rightarrow W^+ W^- \rightarrow q\bar q Q \bar Q$, $q\neq Q$.
The colour factor is 9. If $  q\bar Q (Q\bar q)$ are in a singlet state the
colour factor is 1, if they are in an octet state it is 8.
Therefore the colour  rearrangement probability is 1/9, as is well known.
Similarly, the colour factors for
$e^+e^-\rightarrow q \bar q Q \bar Q$ via electroweak interactions
(EW), $q\neq Q$, 
with $q \bar Q$ in a singlet or octet state are 1 and 8 respectively.

\item
$e^+e^-\rightarrow q\bar q Q \bar Q$ via EW, $q=Q$. 
The full amplitude includes two sets 
of diagrams. The corresponding spinor parts are $A_1$ and $A_2$ 
which are related by the exchange of the momenta,
for example, of the two antiquarks: 

\begin{center}
\begin{picture}(450,110)
\SetScale{.9}
\SetWidth{1.2}
\SetOffset(50,0)

\Text(-45,45)[]{ \Large $A =$}
\Text(0,45)[]{ \Large $A_1$}
%\Photon(30,20)(70,20){5}{4}
%\ArrowLine(70,20)(110,40)
%\ArrowLine(110,0)(70,20)
\ArrowLine(30,20)(70,40)
\ArrowLine(70,0)(30,20)

%\Photon(30,70)(70,70){5}{4}
%\ArrowLine(110,90)(70,70)
%\ArrowLine(70,70)(110,50)
%\Text(115,85)[]{ \large $1$}
%\Text(115,50)[]{ \large $2$}
%\Text(115,35)[]{ \large $3$}
%\Text(115,0)[]{ \large $4$}
\ArrowLine(70,90)(30,70)
\ArrowLine(30,70)(70,50)
\Text(75,85)[]{ \large $1$}
\Text(75,50)[]{ \large $2$}
\Text(75,35)[]{ \large $3$}
\Text(75,0)[]{ \large $4$}

\Text(110,45)[]{ \Large $-$}

\Text(145,45)[]{ \Large $A_2$}
%\Photon(190,20)(230,20){5}{4}
%\ArrowLine(230,20)(270,50)
%\ArrowLine(270,0)(230,20)
\ArrowLine(190,20)(230,50)
\ArrowLine(230,0)(190,20)

%\Photon(190,70)(230,70){5}{4}
%\ArrowLine(270,90)(230,70)
%\ArrowLine(230,70)(270,40)
%\Text(265,85)[]{ \large $1$}
%\Text(265,50)[]{ \large $2$}
%\Text(265,35)[]{ \large $3$}
%\Text(265,0)[]{ \large $4$}
\ArrowLine(230,90)(190,70)
\ArrowLine(190,70)(230,40)
\Text(225,85)[]{ \large $1$}
\Text(225,50)[]{ \large $2$}
\Text(225,35)[]{ \large $3$}
\Text(225,0)[]{ \large $4$}
\end{picture}
\end{center}
If we require the  $(12)$--pair to form a colour singlet, the colour factors 
for the modulus squared of $A_1$ and $A_2$  are 9 and 1
respectively,  and the colour factor 
for the interference between $A_1$ and $A_2$ is 3.
The roles of $A_1$ and $A_2$ are obviously interchanged if instead we
require the $(13)$--pair to form a colour  singlet.

\item
$e^+e^-\rightarrow q\bar q Q \bar Q$ via QCD, $q\neq Q$.
For the   ${\cal O} (\alpha^2 \alpha_s^2)$ tree--level diagrams
the $q\bar q$ and $Q\bar Q$ pairs are always in the octet state.
The colour factor is 2. When $q\bar Q$ form a singlet the colour factor
is 16/9, and when they form an octet the colour factor is 2/9.
In this case the probability of finding the $q\bar Q$ pair
in a singlet state is 8/9.

\item
$e^+e^-\rightarrow q\bar q Q \bar Q$ via QCD, $q=Q$.
The corresponding amplitude includes two sets of four diagrams (only
one is shown for simplicity)
which are related by the exchange of the momenta,
for example, of the two antiquarks: 

\begin{center}
\begin{picture}(450,110)
\SetScale{.9}
\SetWidth{1.2}
\SetOffset(80,0)

\Text(-45,45)[]{ \Large $A =$}
\Text(-5,45)[]{ \Large $A_1$}
%\Photon(30,45)(70,45){5}{4}
%\ArrowLine(70,45)(70,90)
%\ArrowLine(70,90)(110,90)
%\ArrowLine(70,0)(70,45)
%\ArrowLine(110,0)(70,0)
%\Gluon(70,55)(100,55){5}{3}
%\ArrowLine(100,55)(100,75)
%\ArrowLine(100,75)(120,75)
%\ArrowLine(100,35)(100,55)
%\ArrowLine(120,35)(100,35)
%\Text(122,84)[]{ \large $1$}
%\Text(122,69)[]{ \large $2$}
%\Text(122,32)[]{ \large $3$}
%\Text(122,2)[]{ \large $4$}
\ArrowLine(20,45)(20,90)
\ArrowLine(20,90)(60,90)
\ArrowLine(20,0)(20,45)
\ArrowLine(60,0)(20,0)
\Gluon(20,55)(50,55){5}{3}
\ArrowLine(50,55)(50,75)
\ArrowLine(50,75)(70,75)
\ArrowLine(50,35)(50,55)
\ArrowLine(70,35)(50,35)
\Text(72,84)[]{ \large $1$}
\Text(72,69)[]{ \large $2$}
\Text(72,32)[]{ \large $3$}
\Text(72,2)[]{ \large $4$}

\Text(100,45)[]{ \Large $-$}
\Text(130,45)[]{ \Large $A_2$}
%\Photon(180,45)(220,45){5}{4}
\ArrowLine(170,45)(170,90)
\ArrowLine(170,90)(210,90)
\ArrowLine(170,0)(170,45)
\ArrowLine(210,0)(170,0)
\Gluon(170,55)(200,55){5}{3}
\ArrowLine(200,55)(200,75)
\ArrowLine(200,75)(220,75)
\ArrowLine(200,35)(200,55)
\ArrowLine(220,35)(200,35)
\Text(210,84)[]{ \large $1$}
\Text(210,69)[]{ \large $2$}
\Text(210,32)[]{ \large $4$}
\Text(210,2)[]{ \large $3$}

\end{picture}
\end{center}
If we require the  $(14)$ pair to form a colour singlet then 
only $A_2$ contributes and the corresponding colour factor 
is 16/9. The roles of $A_1$ and $A_2$ are obviously interchanged if we
require the  $(13)$ pair to form a colour singlet instead.

\item
$e^+e^- \rightarrow q\bar q gg$.
It is convenient to consider separately the orthogonal contributions which
are symmetric (i.e. proportional to $\{T^a,T^b\} /2$) and 
antisymmetric (i.e. proportional to $[T^a,T^b] /2$) in the gluon colour
indices. The two quarks in the antisymmetric term can only
be in the octet state. The colour factor is then 12.
The colour factor for the symmetric part is 28/3: when the quark pair is in
the singlet state the colour factor is 8/3, and  when the pair is in the
octet state the colour factor is 20/3.  Therefore in kinematic configurations
for which the symmetric and antisymmetric amplitudes are approximately
equal, the colour  singlet pair configurations are suppressed by a
factor $\frac{1}{N_c^2-1}=\frac{1}{8}$.
In particular, configurations which correspond to colour singlet
pairs in opposite hemispheres, which form the background to the
$W^+W^-$ `short string' signal, are colour suppressed.
\end{itemize}

This last result highlights an important difference
in principle between the treatment
of rapidity gap events in the four--quark processes 
(\ref{proc3})--(\ref{proc4}) and the
double gluon bremsstrahlung process (\ref{proc2}).
In the former case two singlet pairs are produced at
 leading order in $\frac{1}{N_c^2}$.
 In the latter case one is dealing with a colour suppressed
phenomenon where there is an important difference
between the probabilistic exclusive  (e.g. Lund string) picture
 and the interference inclusive (perturbative QCD)
treatment of the final state particle distributions
\cite{gamw,inter2,pert}. Within the inclusive perturbative QCD scenario,
in which the interference between radiative amplitudes is fully taken into
account, the
presence of the different octet and singlet colour states for
 the $gg$ system most likely
leads to  small (i.e. of order
$\frac{1}{N_c^2}$) anisotropies in the particle distributions rather
than to the appearance of distinctive rapidity gap events with small
$O(\frac{1}{N_c^2})$ probability. In other words, the colour singlet
$gg$ configuration would correspond to an `accidental
singlet' \cite{inter2} with no obvious probabilistic
interpretation: additional gluon radiation would
`smear out' the colour structure on an event--by--event basis.
 On the other hand, one might  expect
that these colour singlet pair events {\it would}
manifest themselves on an event--by--event basis  when the invariant
masses of the dijet systems are sufficiently small,
in which case  non--perturbative
dynamics (like  string formation in the Lund model) would be
the dominant effect. This would  correspond to
 two ultra--relativistic small colour dipoles receding from each other
with high velocities, too far apart to interact. In the present study,
analogously to Refs.~\cite{tr,fme},  we will adopt the latter
(probabilistic exclusive)
scenario for the final state structure of the $q\bar qgg$ events.\footnote{For
a recent comprehensive discussion of the above colour connection
phenomena in processes such as $e^+e^-\to q \bar{q} g \ldots g$
see Ref.~\cite{fgh}.}

\section{Results}

In this section we quantify the relative importance of the 
signal and background contributions in the production of rapidity
gap events at LEP2. 
For the numerical calculations we take
$\alpha_{em}= 1/128$ and  $\sin^2\theta_W=0.23$, while
for the $Z^0$ and $W^\pm$  boson masses and widths we use the values
$M_{Z^0}=91.1$~GeV, $M_{W}=80.0$~GeV, and 
$\Gamma_{Z^0}=2.5$~GeV, $\Gamma_{W}=2.2$~GeV.
% respectively, computing
%the $W$ mass using the Standard Model
% relation $M_{W}=M_{Z^0}\cos\theta_W\approx 
%80$~GeV, and with $\Gamma_{W}=2.2$~GeV.
The final-state quarks are taken to be massless.
The strong coupling constant 
$\alpha_s$ is computed
at  the two--loop level, with five active quark flavours, at a scale equal to 
the collider energy $\sqrt s$ and
with $\Lambda_{QCD}=190$ MeV.\footnote{In principle, the natural scale for
the strong coupling constant $\alpha_s$ is the characteristic momentum
transfer, in which case the effective value of $\alpha_s$ should be somewhat
larger than $\alpha_s(\sqrt s)$  
due to the lower values of the momenta. However, in what
follows we shall treat $\alpha_s$ as a constant ($\approx0.105$).}
The analysis is performed at the parton level, neglecting
the effects of hadronization. 

In order to be consistent with the experimental procedure of selecting 
4--jet final states, we need to adopt a jet--finding algorithm.
We choose the `Durham' scheme \cite{DURHAM}, which uses
the clustering variable 
\be\label{DURHAM}
y^D_{ij} = {{2\min (E^2_i, E^2_j)(1-\cos\theta_{ij})}
\over{s}},
\ee
where $E_i$ and $E_j$ are the energies of the $i$--th and $j$--th
particle (with four--momenta $p_i$ and $p_j$), respectively, and
$\theta_{ij}$ their relative angle.
A 4--jet event is then defined by requiring that all $y_{ij}$'s (obtained
from all  possible permutations of $i$ and $j$) satisfy
the condition $y_{ij}\ge y_{cut}^D=0.0015$. None of  the main 
features of our results  depend significantly on the jet--clustering 
procedure and/or on the exact value of $y_{cut}$. 

Our results are presented in  Table~I and in  Figs.~5--8.
In Table~I (upper section) 
we present the total cross sections for the four processes
\eewwqqqq, \eeqqgg, \eeqqqq\ (via $g$--propagators), and 
\eeqqqq\ (via $\gamma,Z^0$--propagators), for $y_{cut}^D=0.0015$
and $\sqrt s=180$~GeV. These correspond to the `fully coloured'
 matrix elements in that both colour singlet and octet 
contributions are included.\footnote{All the results for process
(\ref{proc1})
presented here neglect  gluon radiation and cascading effects and, therefore, 
are intended  only for illustrative purposes -- see the discussion at
the end of this section.} 
In computing these cross sections we have summed (averaged) over
the final (initial) state helicities and summed over all 
possible combinations of quark flavours in the final states.
%In contrast to the situation at LEP1,  where quark mass effects can be
% important, at LEP2 quark masses in 4--jet events 
%can be safely neglected 
%%\cite{noiPL,noiNP,noiPr}. 
%\cite{noiNP,noiPr}. 
For the processes under consideration, quark masses can safely be neglected.

Since $2M_{W}<\sqrt s<2M_{Z^0}$ the cross section for
the four-quark process mediated by the double $W^\pm$
resonance (i.e. \eewwqqqq)
is much larger than the EW \eeqqqq\, which includes diagrams with 
a double $Z^0$ resonance.\footnote{Increasing or decreasing the
value of $\sqrt s$ will,  of course, lead to a different relative
weighting of these processes. Our choice here of $\sqrt s =  180$~GeV 
corresponds to maximising the number of $W^+W^-$ events while
remaining below the nominal $Z^0Z^0$ threshold.}
Even though the two--gluon QCD process \eeqqgg\ 
is produced  via $\gamma^*,Z^0$ $s$--channel exchange, giving a
$1/s$ suppression,
the corresponding 4--jet cross section is comparable to that for
\eewwqqqq. 
Finally, the QCD  \eeqqqq\ cross section is comparable to that
for \eeqqqq\ via EW interactions,\footnote{This is true for the 
particular value
of $y_{cut}$ chosen here. The absolute rates are quite sensitive
to the $y_{cut}$ value, but much of this sensitivity disappears when
further kinematical cuts and colour-singlet selections are applied,
see below.}
and roughly  an order
of magnitude smaller  than the cross section for 
processes (\ref{proc1}) and (\ref{proc2}),
see also Ref.~\cite{fjo}. 

Figure~5 shows the corresponding differential distribution $d\sigma/dE_j$
of the energy of a single jet in the four processes, for the 
above values of $\sqrt s$ and $y_{cut}^D$.
The symmetry of the matrix elements under interchange
of the (anti)quark labels means that the energy distributions
are identical for all (massless) partons  in the 
\eewwqqqq\ and (QCD and EW) \eeqqqq\ processes.
In contrast, the distributions are {\it different} for the quarks and gluons
in \eeqqgg\ production. The infrared singularities 
in gluon bremsstrahlung off quarks in $q\bar qgg$ production
lead to a softer energy distribution for the gluons. 
 Another notable feature in Fig.~5 is the peak
around $E_j = 1/4\sqrt s$ for \eewwqqqq\ events,  due to
the fact that for $\sqrt s\approx2M_{W}$ the energy of the quarks
from $W^\pm$ decays is approximately $M_{W}/2\approx
1/4\sqrt s$. The spread of the distribution in Fig.~5 about this value
is due mainly to the kinetic energy of the $W^\pm$ bosons at this
above--threshold collision energy.
A similar effect  is observed in the EW \eeqqqq\ distribution, due
to the double $Z^0$--resonance contribution.
 Here, however,  the additional  presence (see Fig.~3a) of
$\gamma^*\ar q\bar q$ propagators with soft singularities tends 
to shift the energy to small values.
 Finally, the energy distribution of QCD \eeqqqq\ events 
peaks at the edges of the range, since in these events we have quarks both from
$Z^{0*}$ decays with virtuality $\sqrt s$  (and thus $E_j\approx\sqrt s/2$) and
from soft virtual gluons (i.e. with $E_j\ar 0$). 

In order to study colour rearrangement in 4--jet
production at LEP2 we are
interested in  large thrust events in which
all jets have approximately the same energy \cite{model1}.
In Fig.~6 we therefore show  the differential  thrust distribution $d\sigma/dT$
 for $y_{cut}^D=0.0015$,
and with an `equal energy' cut, viz.
 $|E_j-{\sqrt s}/{4}|<10$~GeV for all final--state jets.
All colour combinations are again included. 
The distributions are quite flat for processes (\ref{proc1}), 
(\ref{proc3}) and (\ref{proc4})  for $0.7\Ord T\Ord 0.9 - 0.95$,
decreasing as $T\ar 1$ due to the finite $y_{cut}^D$ cut.
For the processes which are mediated by two virtual massive
bosons ($W^+W^-$ or $Z^0Z^0$) the jets are roughly isotropic
in phase space since the collision energy is not far
from threshold. 
For the contribution to (\ref{proc4}) in which the $q\bar q Q\bar Q$
are produced via intermediate photons, we expect large thrust events when 
$\gamma^*\ar q\bar q$ propagators have small invariant masses (i.e.
collinear $q$ and $\bar q$). This happens 
in $\gamma^*\gamma^*$ back--to--back events from the  diagrams in Fig.~3b,
in $3+1$ back--to--back configurations (i.e. with three collinear jets  
in one hemisphere) in $\gamma^* Z^{0*}$  events from Fig.~3b, and in
the QCD--like diagrams of Fig.~3a. Since these latter graphs 
apply also to 
$g^*\ar q\bar q$, the same comments hold for process (\ref{proc3}).
In contrast, the thrust distribution for the QCD process
(\ref{proc2}) increases significantly at large $T$,  due to the collinear
singularities of the matrix element. 
For example, a single gluon bremsstrahlung
 from each of the two quarks gives a back--to--back $2+2$ configuration,
 while a gluon
splitting into a $gg$--pair via a triple gluon vertex 
gives a  $3+1$ configuration. 

Since our aim is to   predict the relative contributions
of processes (\ref{proc1})--(\ref{proc4}) to 4--jet events with large
rapidity gaps and to study
colour rearrangement effects, and since these latter occur
between parton pairs in  colour singlet
states (i.e. $q\bar q$, $q\bar Q$, $Q\bar Q$, $Q\bar q$ or $gg$) 
sufficiently close in phase space, we show in Fig.~7  the distribution
 in the angular separation
 $\cos\theta_S$  between all possible parton pairs
which can give such colour singlet  configurations.
In process (\ref{proc1}) singlet states can occur both in 
$[q\bar q][Q\bar Q]$      (both quark and antiquark from 
the same $W$) and in
$[q\bar Q][Q\bar q]$ (quark and antiquark from different $W$'s)
combinations. 
For the former, the distribution is peaked in the back--to--back direction
(the $W$'s are only slowly moving at this collision energy)
whereas the latter distribution is much flatter. Near  $\cos\theta_S =  1$
the distribution is suppressed by the $y_{cut}^D$ cut.
 For the QCD $q \bar q gg$ process
(\ref{proc2}) only $q\bar q$ and $gg$ pairs can give 
colour singlet states. The peak in the distributions at $\cos\theta_S\ar -1$ 
for both of these combinations reflects the dominance
of the  back--to--back configuration of the $q\bar q$ pair in the 
centre--of-mass frame, with gluons preferentially emitted
along the quark and antiquark directions.   
For $\cos\theta_S\ar 1$ we see a small
increase for the $gg$ distribution
due to the collinear singularity in the $g^*$ propagator splitting
into $gg$--pairs. 
The distribution in
$\cos\theta_S$ for the QCD $q \bar q Q \bar Q$ process 
(\ref{proc3})\footnote{Note that only the $[q\bar Q][Q\bar q]$ 
combination can give a colour singlet here.}
is also strongly peaked in the back--to--back direction.
Finally, the $q\bar q$ distribution of the  EW
 $q \bar q Q \bar Q$ process (\ref{proc4}) has almost
the same features as that  of process (\ref{proc1}), apart
from the stronger peaking peak at $\cos\theta_S\ar -1$ which
reflects the fact that the $Z^0Z^0$ pairs are produced almost at rest 
at this collision energy.
As for \eewwqqqq, the distribution  for $q\bar Q$ is again
quite flat. 
% (the value at $\cos\theta_S\ar -1$ is $\approx0.07$ pb
%whereas for $q\bar q$ is at $\approx0.35$ pb), and for the same reasons
%as for process (\ref{proc1}).

The total cross sections corresponding to Figs.~6--7
are listed  in Table~I (central section), 
for the same choice of cuts and colour factors. 
Because of the jet energy dependence of processes    
(\ref{proc2})--(\ref{proc3}) (see Fig.~5), 
their overall rates are much smaller compared to the other two
which are 
`energy  resonant' in the region $E_j \approx \sqrt s/4$. 

The final plot, Fig.~8, shows the thrust dependence of those events
which can give rapidity gap configurations. Here
only the colour singlet part of the matrix elements
has been computed and particles forming singlet states have been
required to be close in phase space (i.e. $\cos\theta_S>0$).
 Note that for processes (\ref{proc1}) and (\ref{proc4})
  both $[q\bar q][Q\bar Q]$  and $[q\bar Q][Q\bar q]$ colour singlet
configurations are included.
It is interesting to note that requiring only positive cosines of $\theta_S$
and singlet colour factors in the matrix elements almost completely
eliminates the collinear singular configurations 
of the QCD 4--jet events which were apparent in Fig.~6. 
The effect is clearly visible
as a  decrease in the thrust dependence
of these two processes at large $T$. 

In Table~I (lower section) 
we list  the singlet total cross sections of the four processes,
for $y_{cut}^D=0.0015$, $|E_j-{\sqrt s}/{4}|<10$~GeV and 
$\cos\theta_S>0$, i.e. the integrated distributions of
Fig.~8. 
There is an obvious hierarchy of cross sections:
$q \bar q Q \bar Q (W^+W^-) \; \gg \; q \bar q Q \bar Q (EW) \; \gg
 \; q \bar q Q \bar Q, q \bar q gg (QCD)$, the difference 
 in each case being approximately one order of magnitude.
  If one assumes  an integrated
luminosity ${\cal L}=500$~pb$^{-1}$, only
\eewwqqqq\ production will   give a sizeable
number of colour reconnection events at LEP2.
 
We end this section with a word of caution. 
The estimate of the number of rapidity gap events for  process (\ref{proc1})
based on the above four--parton results
almost certainly
overestimates the size of the colour rearrangement effects.
 It corresponds to the so--called `instantaneous
reconnection scenario' (analogous to Ref.~\cite{lab}), in which
 the colour singlet  $(q\bar Q)$ and $(Q\bar q)$ states
are instantaneously formed and allowed to radiate perturbative
gluons. It was first shown in Ref.~\cite{inter2} that  the finite
$W$ width leads to a  space--time separation between the $W$'s,
and a consequent additional suppression (by at least an order
of magnitude) of the
recoupling effects. Therefore the rate of `short string'
or rapidity gap production
corresponding to processes (\ref{proc1}) and (\ref{proc4}) 
may well be  of the same
order. However the low overall event rate will surely make these
difficult to identify experimentally.
 
\section{Conclusions}

One of the most important uncertainties in the direct reconstruction method of
determining the $W$ mass 
from  $q\bar q Q\bar Q$ events at LEP2 is related to the
relatively unknown QCD interconnection effects 
\cite{inter2,model1,model2,jekg}. To
obtain  information about the size of these effects it was proposed
in Ref.~\cite{model1} to study the `short string' states signal in 
$W^+W^-$ hadronic events in which colour singlet states
are formed from $q \bar Q $ and $Q\bar q$ pairs with small phase--space
separation. However, such configurations can also be
generated by the conventional background $e^+e^-\ar 4$~parton events in 
kinematic configurations corresponding to rapidity gap events
\cite{trett,tr,fjo,fme}. In this paper we have presented quantitative
estimates of the expected rate from these background processes,
including not only the QCD four--parton processes but also the
non--resonant  electroweak $q \bar q Q \bar Q$ processes.
In each case we have computed the cross sections corresponding
to two pairs of colour  singlet partons with modest angular
separation and approximately equal parton (jet) energies.
We have shown that in this configuration the QCD processes
are heavily suppressed, but that  electroweak $q \bar q Q \bar Q$ 
production may be comparable in rate to the expected signal
from $W^+W^-$ production.  

Finally, we note that aside from their importance to the 
$M_W$ measurement at LEP2, rapidity gap events
in \epem\ annihilation are interesting in their own right as a new laboratory
for studying the dynamics of hadron production, see for example 
Refs.~\cite{model1,tr,fgh}. One of
the important issues here is the correspondence between the
colour and particle flow dynamics, see for example Ref.~\cite{pert}.
It would be straightforward to extend our calculations to other
collider energies and kinematical configurations in order to 
study such effects.

\section*{Acknowledgements}

This work is supported 
in part by the Ministero dell' Universit\`a e della Ricerca Scientifica,
and   the EU Programme
`Human Capital and Mobility', Network `Physics at High Energy
Colliders', contract CHRX-CT93-0357 (DG 12 COMA).
VAK, SM and WJS are grateful to the UK PPARC for support. We thank 
G.~Gustafson and T.~Sj\"ostrand for valuable discussions.

\vfill
\newpage
\thispagestyle{empty}

\section*{Table Caption}

\begin{itemize}

\item[{[1]}] Total cross sections for the four processes:
                 {\it (i)}   \eewwqqqq,      
                 {\it (ii)}  \eeqqgg,    
                 {\it (iii)} \eeqqqq\ (QCD),
                 {\it (iv)}  \eeqqqq\ (EW),
                 at $\sqrt s=180$~GeV, for $y_{cut}^D=0.0015$:
                 with complete matrix elements and
                 no additional kinematical cuts  (upper section);
                  with complete matrix elements  and with the additional
                 cut $|E_j-{\sqrt s}/{4}|<10$~GeV (central section);
                 and with only the `singlet' component of the matrix elements 
                 and with  the cuts
                 $|E_j-{\sqrt s}/{4}|<10$~GeV and $\cos\theta_S>0$ (lower
                 section).
 Note that for processes (\ref{proc1}) and (\ref{proc4})
  both $[q\bar q][Q\bar Q]$  and $[q\bar Q][Q\bar q]$ colour singlet
configurations are included.

\end{itemize}

\vfill
\newpage
\thispagestyle{empty}

\section*{Figure Captions}

\begin{itemize}

\item[{[1]}]  Feynman diagrams contributing in lowest order to
                  $e^+e^-\rightarrow W^+W^-\ar q\bar q Q\bar Q$.
                  A wavy line represents a $\gamma$, a $Z^0$ or a $W^\pm$, as
                  appropriate.
                  External lines are identified by their indices
                  as given in the text.

\item[{[2]}]  Feynman diagrams contributing in lowest order to
                  $e^+e^-\rightarrow q\bar q gg$.
                  A wavy line represents a $\gamma$ or a $Z^0$ while the 
                  helical line represents a $g$.
                  External lines are identified by their indices
                  as given in the text.

\item[{[3]}] Feynman diagrams contributing in lowest order to
                 $e^+e^-\rightarrow q\bar q Q\bar Q$ via QCD 
                 (a) and via EW (a and b).
                 If the two quark 
                 pairs have different flavour only the first four diagrams
                 in Fig.~3a and the first two in Fig.~3b contribute.
                 A wavy line represents a $\gamma$ or a $Z^0$ 
                 while a jagged line
                 represents a $g$, a $\gamma$ or a $Z^0$, as appropriate.
                 External lines are identified by their indices as 
                 given in the text.

\item[{[4]}] Four--parton mechanisms for generating  rapidity gap
                 events in processes (\ref{proc1})--(\ref{proc4}).
                 The dashed lines indicate that the produced
                 partons are in colour singlet states: (a) a quark--antiquark
                 jet pair and a two--gluon jet pair; (b) two final
                 quark--antiquark jet pairs produced via QCD interactions; (c)
                 and (d) two final
                 quark--antiquark jet pairs produced via EW interactions.

\item[{[5]}] Differential distribution $d\sigma/dE_j$
                 in  the energy of the jet(s) for the four processes:
                 {\it (i)}   \eewwqqqq\       (continuous line),
                 {\it (ii)}  \eeqqgg\     (dashed line),
                 {\it (iii)} \eeqqqq\ (QCD) (dotted line), and
                 {\it (iv)}  \eeqqqq\ (EW)  (chain--dashed line), 
                 at $\sqrt s=180$~GeV, for $y_{cut}^D=0.0015$.
                 The complete matrix elements are used.

\item[{[6]}] Differential distribution $d\sigma/dT$
                 in thrust for the four processes:
                 {\it (i)}   \eewwqqqq\       (continuous line),
                 {\it (ii)}  \eeqqgg\     (dashed line),
                 {\it (iii)} \eeqqqq\ (QCD) (dotted line), and
                 {\it (iv)}  \eeqqqq\ (EW)  (chain--dashed line), 
                 at $\sqrt s=180$~GeV, for $y_{cut}^D=0.0015$
                 and $|E_j-{\sqrt s}/{4}|<10$~GeV.
                 The complete matrix elements are used.
                  
\end{itemize}

\vfill
\newpage
\thispagestyle{empty}

\begin{itemize}

\item[{[7]}] Differential distribution $d\sigma/d\cos\theta_S$
                 in the cosine of the `singlet angle'
                 for the four processes:
                 {\it (i)}   \eewwqqqq\       (continuous line),
                 {\it (ii)}  \eeqqgg\     (dashed line),
                 {\it (iii)} \eeqqqq\ (QCD) (dotted line), and
                 {\it (iv)}  \eeqqqq\ (EW)  (chain--dashed line), 
                 at $\sqrt s=180$~GeV, for $y_{cut}^D=0.0015$
                 and $|E_j-{\sqrt s}/{4}|<10$~GeV.
                 The complete matrix elements are used.

\item[{[8]}] Differential distribution $d\sigma/dT$
                 in thrust for the four processes:
                 {\it (i)}   \eewwqqqq\       (continuous line),
                 {\it (ii)}  \eeqqgg\     (dashed line),
                 {\it (iii)} \eeqqqq\ (QCD) (dotted line), and
                 {\it (iv)}  \eeqqqq\ (EW)  (chain--dashed line), 
                 at $\sqrt s=180$~GeV, for $y_{cut}^D=0.0015$,
                 $|E_j-{\sqrt s}/{4}|<10$~GeV and $\cos\theta_S>0$.
                 Only singlet components of the matrix elements were used.
 Note that for processes (\ref{proc1}) and (\ref{proc4})
  both $[q\bar q][Q\bar Q]$  and $[q\bar Q][Q\bar q]$ colour singlet
configurations are included.

\end{itemize}

\vfill
\newpage
\thispagestyle{empty}
\
\vskip8.0cm
\begin{table}%[p]%[htbp]
\begin{center}
\begin{tabular}{|c|c|c|c|}
\hline
\multicolumn{4}{|c|}
{\rule[0cm]{0cm}{0cm}
$\sigma$ (pb)}
\\ \hline
\rule[0cm]{0cm}{0cm}
\wwqqqq\ & \qqgg\ & \qqqq\ (QCD) & \qqqq\ (EW) \\ \hline\hline
\rule[0cm]{0cm}{0cm}
$6.34$ & $4.33$ & $0.240$ & $0.186$  \\ \hline 
\multicolumn{4}{|c|}
{\rule[0cm]{0cm}{0cm}
no kinematical cuts \quad\quad\quad\quad singlet+octet }
 \\ \hline\hline

\rule[0cm]{0cm}{0cm}
$1.32$ & $6.84\times10^{-2}$ & $1.06\times10^{-3}$ & $4.75\times10^{-2}$  
\\ \hline
\multicolumn{4}{|c|}
{\rule[0cm]{0cm}{0cm}
$|E_j-{\sqrt s}/{4}|<10$~GeV \quad\quad\quad\quad singlet+octet }
 \\ \hline\hline

\rule[0cm]{0cm}{0cm}
$2.54\times10^{-2}$ & $5.16\times10^{-5}
$ & $9.44\times10^{-5}$ & $3.01\times10^{-3}$    \\ \hline
\multicolumn{4}{|c|}
{\rule[0cm]{0cm}{0cm}
$|E_j-{\sqrt s}/{4}|<10$~GeV \quad\quad $\cos\theta_S>0$
\quad\quad singlet only}
 \\ \hline
 
% \hline
%
%\multicolumn{4}{|c|}
%{\rule[0cm]{0cm}{0cm}
%$\sqrt s=180$~GeV\quad\quad\quad $y_{cut}^D=0.0015$}
% \\ \hline
 
\multicolumn{4}{c}
{\rule{0cm}{1.0cm}
{\Large Table I}}  \\
\multicolumn{2}{c}
{\rule{0cm}{0cm}}

\end{tabular}
\end{center}
\end{table}

\vfill
\newpage
\thispagestyle{empty}

\begin{figure}[p]~\epsfig{file=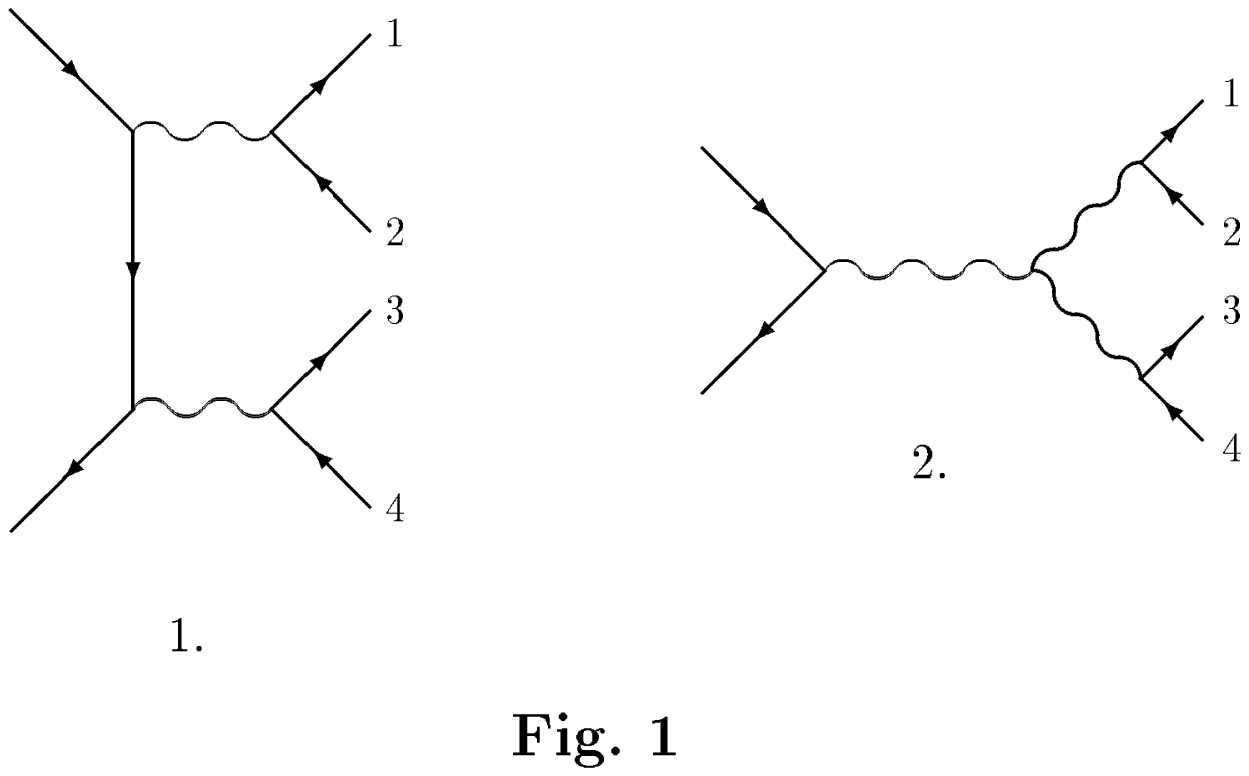,height=22cm}
\end{figure}
\stepcounter{figure}
\vfill
\clearpage
\thispagestyle{empty}

\begin{figure}[p]~\epsfig{file=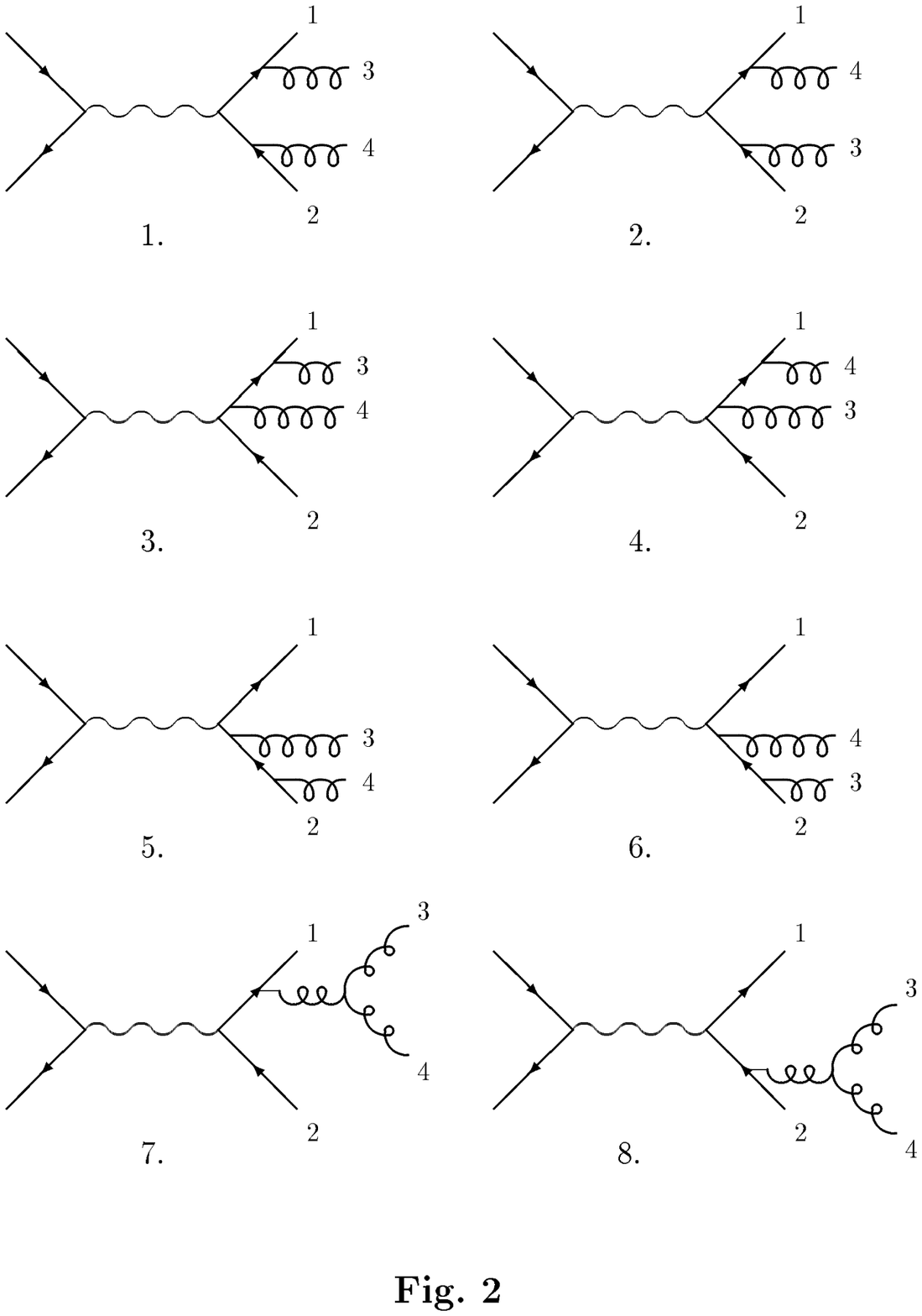,height=22cm}
\end{figure}
\stepcounter{figure}
\vfill
\clearpage
\thispagestyle{empty}

\begin{figure}[p]~\epsfig{file=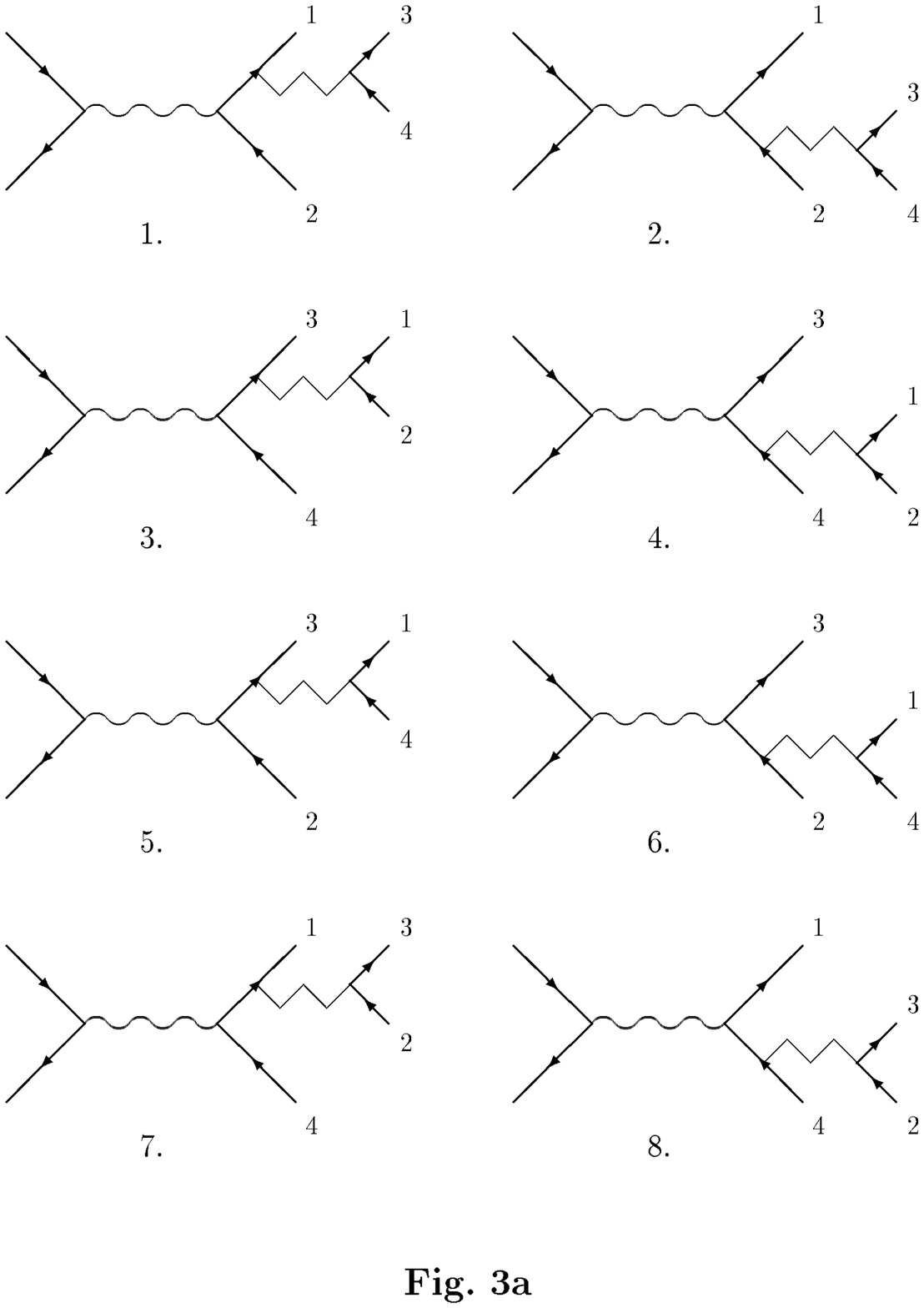,height=22cm}
\end{figure}
\stepcounter{figure}
\vfill
\clearpage
\thispagestyle{empty}

\begin{figure}[p]~\epsfig{file=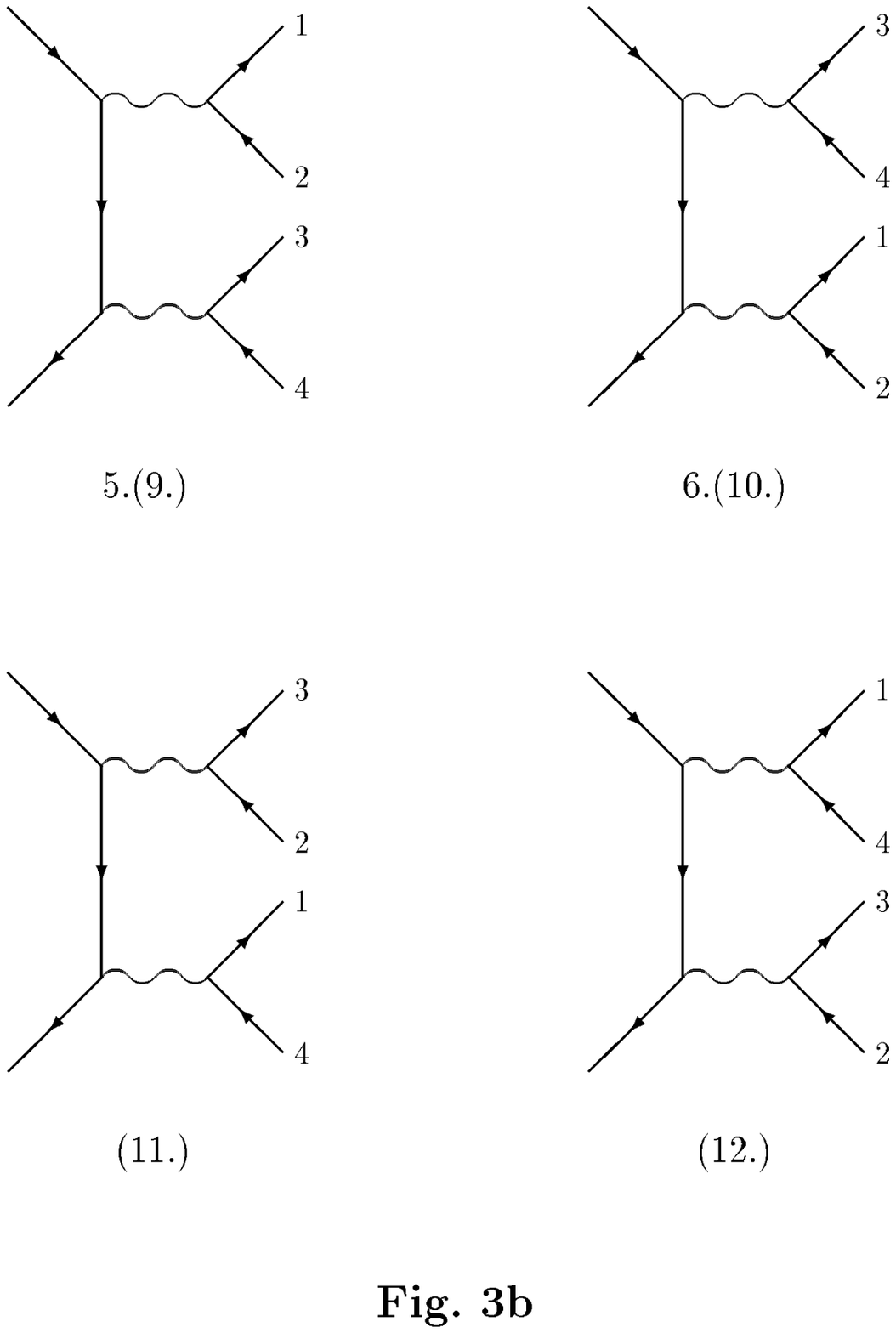,height=22cm}
\end{figure}
\stepcounter{figure}
\vfill
\clearpage
\thispagestyle{empty}

\begin{figure}[p]~\epsfig{file=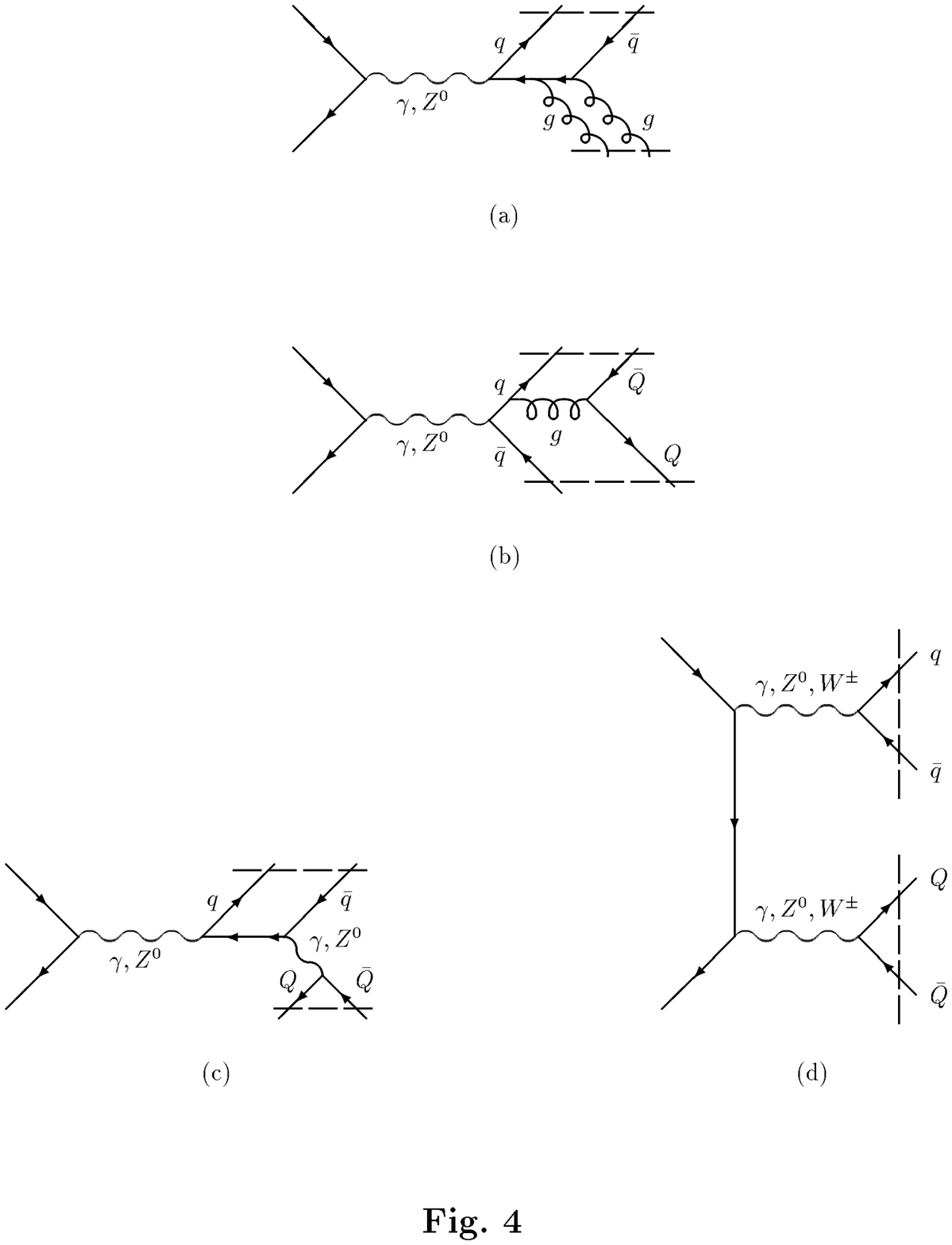,height=22cm}
\end{figure}
\stepcounter{figure}

\vfill
\clearpage
\thispagestyle{empty}
\begin{figure}[p]~\epsfig{file=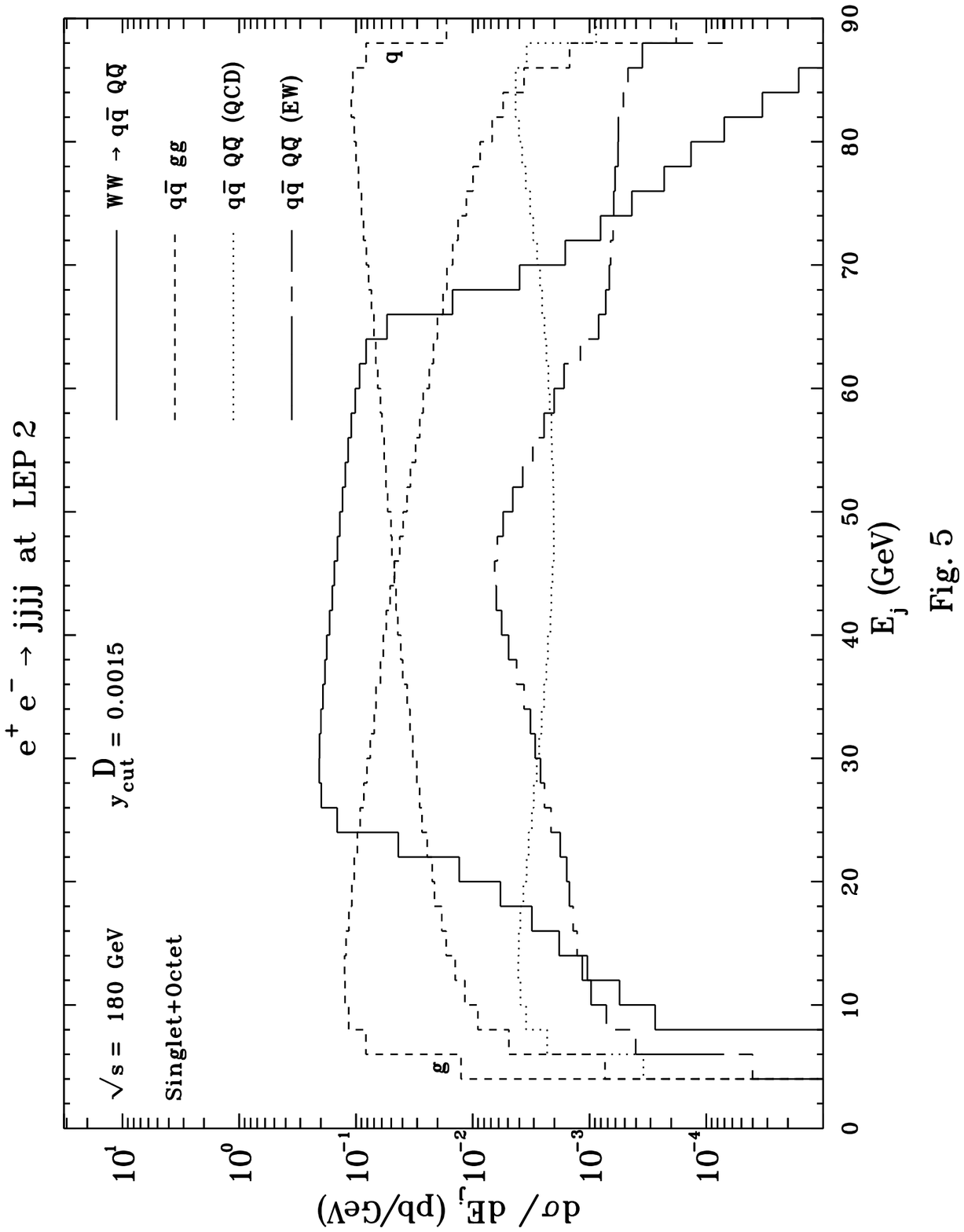,height=22cm}
\end{figure}
\stepcounter{figure}
\vfill
\clearpage
\thispagestyle{empty}

\begin{figure}[p]~\epsfig{file=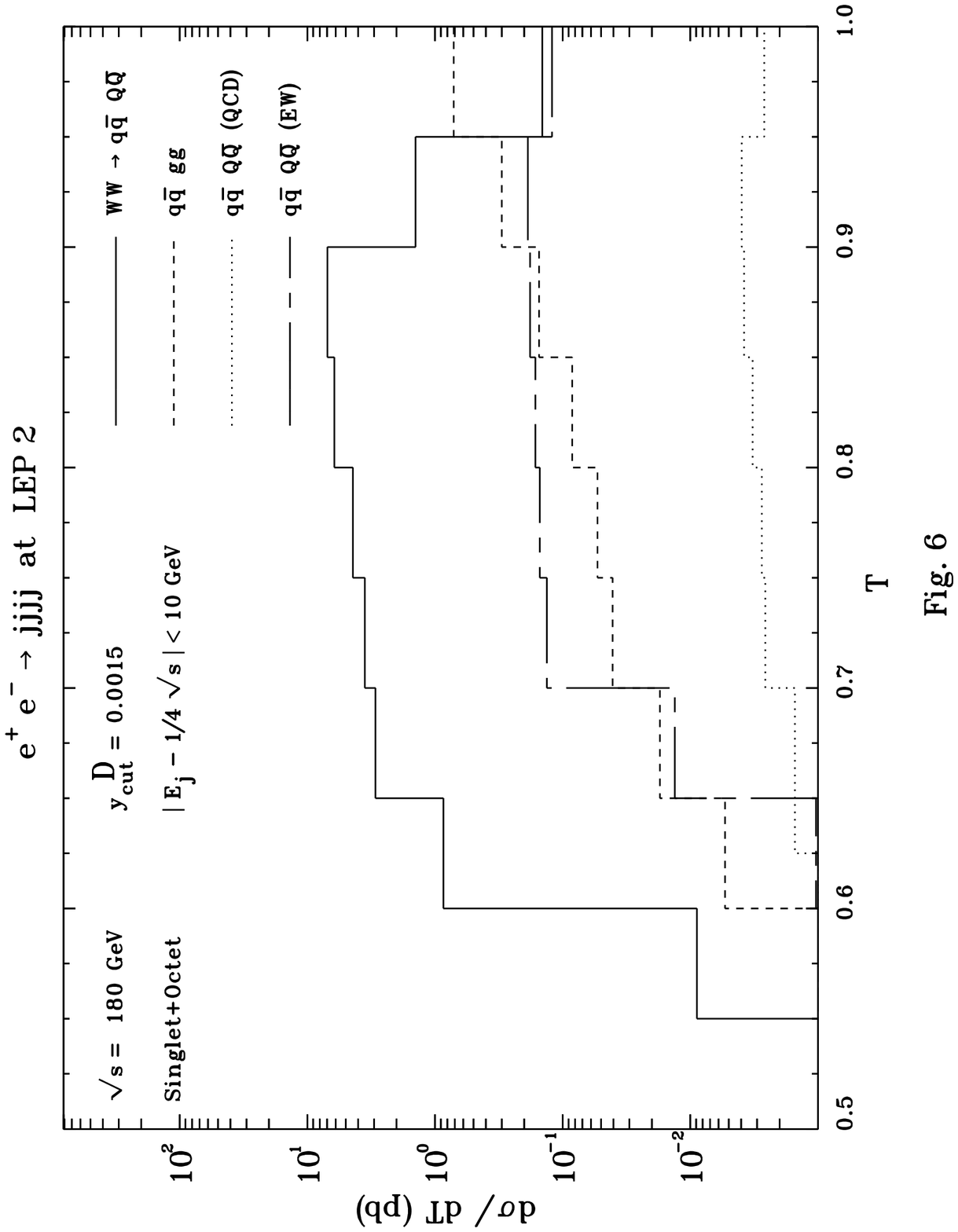,height=22cm}
\end{figure}
\stepcounter{figure}
\vfill
\clearpage
\thispagestyle{empty}

\begin{figure}[p]~\epsfig{file=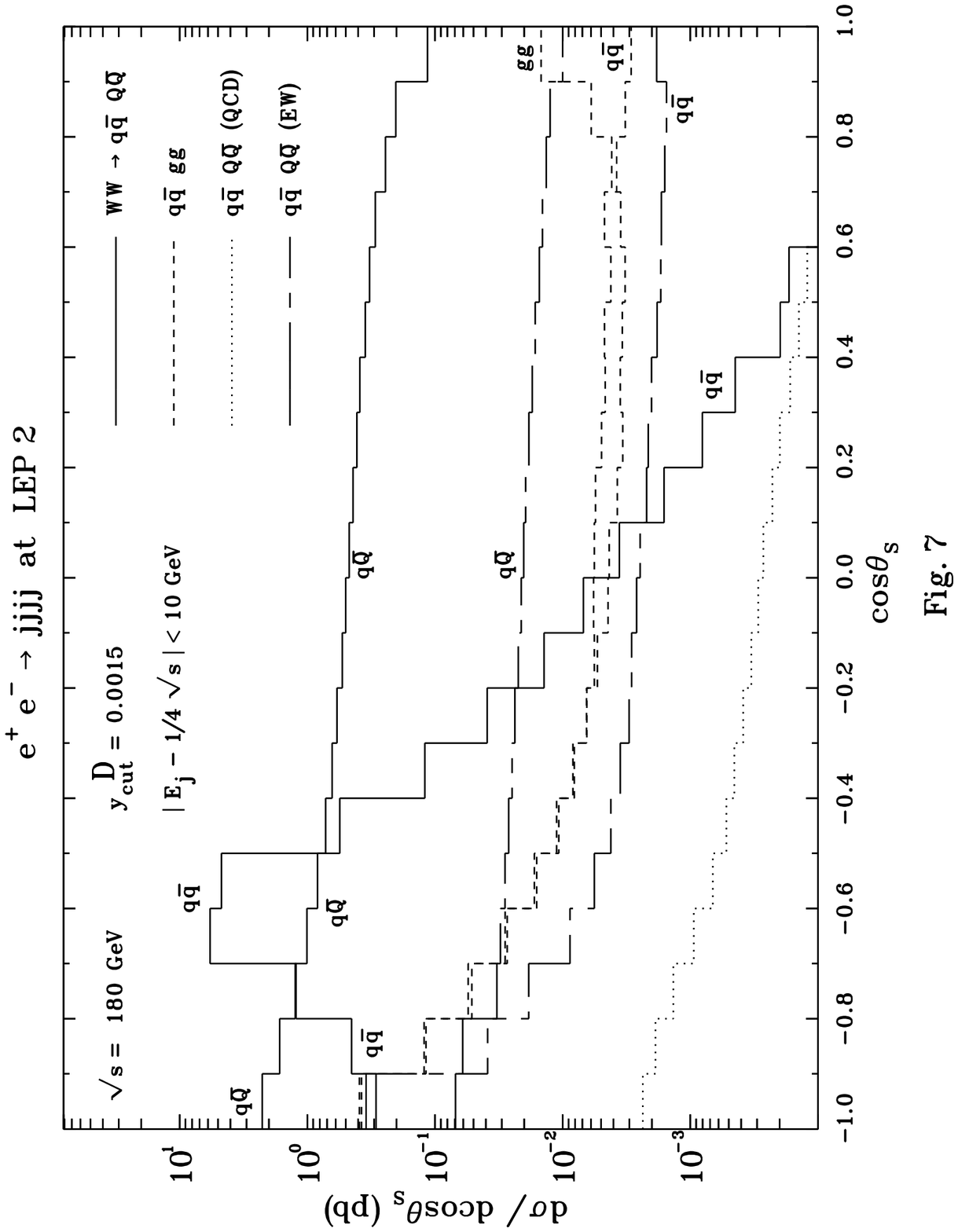,height=22cm}
\end{figure}
\stepcounter{figure}
\vfill
\clearpage
\thispagestyle{empty}

\begin{figure}[p]~\epsfig{file=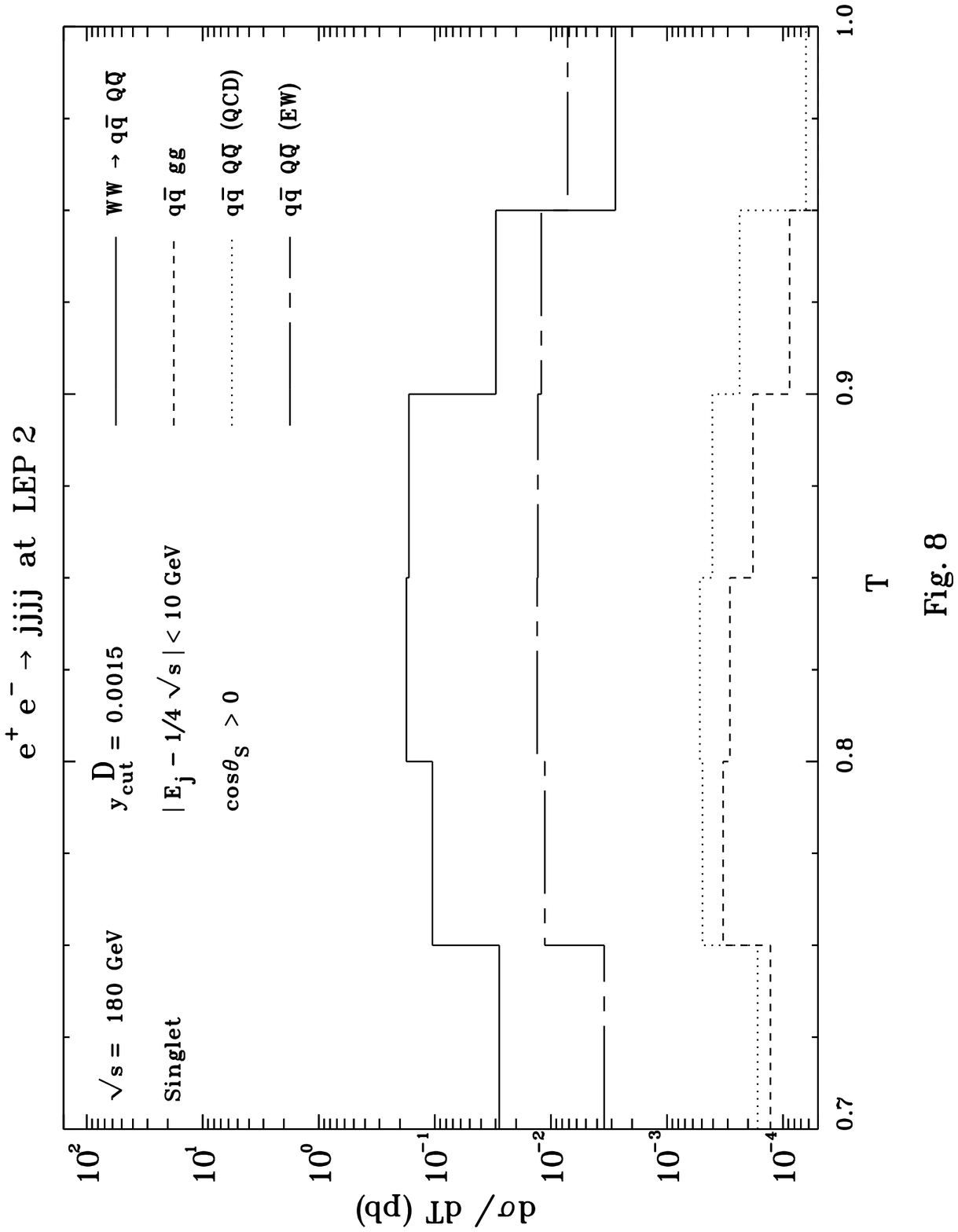,height=22cm}
\end{figure}
\stepcounter{figure}

\vfill

\begin{thebibliography}{99}

\bibitem{accuracy} 
`Determination of the Mass of the W Boson', Z. Kunszt {\it et al.},
in `Physics at LEP2', eds. G. Altarelli, T. Sj\"{o}strand and F. Zwirner,
CERN Report 96-01, Vol.1, p.141-205 (1996). 

\bibitem{gamw} For a recent review, see for example:
               V.A. Khoze in Proc. First Arctic Workshop
               on Future Physics and Accelerators, Saariselsk\"a, Finland,
               August 1994, eds. M. Chaichan,
               K. Huitu and R. Orava, World Scientific, Singapore, 1995, p.458.
 
\bibitem{inter1} V.S. Fadin, V.A. Khoze and A.D. Martin, \pr D49 1994 2247;
 \pl B320 1994 141.

\bibitem{inter2} T. Sj\"ostrand and V.A. Khoze, \zp C62 1994 281;~{\it
                 Phys. Rev. Lett.}~{\bf 72}~(1994)~28.                 

\bibitem{model1} G. Gustafson and J. H\"akkinen, \zp C64 1994 659.

\bibitem{model2} B.R. Webber, `Colour reconnection in HERWIG',
                 talk given at the LEP2 QCD Working Group meeting,
                 CERN, Geneva, Switzerland, May 1995.

\bibitem{jekg} J. Ellis and K. Geiger, \preprint\ CERN--TH/95--283 (1995).

\bibitem{lab} G. Gustafson, U. Petterson and P.M. Zerwas, \pl B209
              1988 90.

\bibitem{pert} Yu.L. Dokshitzer, V.A. Khoze, A.H. Mueller and S.I.
Troyan, `Basics of Perturbative QCD', ed. J. Tran
Thanh Van, Editions Frontieres, Gif--sur--Yvette, 1991;~\rmp 60 1988 373.

\bibitem{swe} B. Andersson, G. Gustafson, G. Ingelman and T.
Sj\"ostrand, \prep 97 1983 31.

\bibitem{fong} Ya.I. Azimov, Yu.L. Dokshitzer, V.A. Khoze and S.I.
Troyan, \pl B165 1985 147.

\bibitem{long} Yu.L. Dokshitzer, V.A. Khoze and S.I. Troyan, \sjnp 50
1990 505.

\bibitem{twe} L. L\"onnblad and T. Sj\"ostrand, \pl B351 1995 293.

\bibitem{trett} J. Randa, \pr D21 1980 1795.

\bibitem{tr} J.D. Bjorken, S.J. Brodsky and H.J. Lu, \pl B286 1992
153.

\bibitem{fjo} H.J. Lu, S.J. Brodsky and V.A. Khoze, \pl B312 1993 215.

\bibitem{fme} J. Ellis and D.A. Ross,  \zp C70 1996 115.

\bibitem{metodo}  A. Ballestrero and E. Maina, \pl B350 1995 225.

%\bibitem{eezh}  A. Ballestrero, E. Maina and S. Moretti, \pl B335 1994 460.

\bibitem{noiPL} A. Ballestrero, E. Maina and S. Moretti, 
{\it Phys. Lett.} {\bf B294} (1992) 425.

\bibitem{noiNP} A. Ballestrero, E. Maina and S. Moretti, 
{\it Nucl. Phys.} {\bf B415} (1994) 265.

\bibitem{noiPr} A. Ballestrero, E. Maina and S. Moretti, 
{\it University of Torino preprint} DFTT 14/94,
                (1994), to be 
                published in  Proc. XXIXth Rencontres de 
                Moriond, M\'eribel, Savoie, France, March 1994.

\bibitem{ks}  R. Kleiss and W.J. Stirling, 
{\it Nucl. Phys.} {\bf B262} (1985) 235.

%\bibitem{review} S. Bethke, Z. Kunszt, D.E. Soper and W.J. Stirling, 
%\np B370 1992
%               310.

\bibitem{fgh} C. Friberg, G. Gustafson and J. H\"akkinen, 
{\it University of Lund preprint}  LU-TP-96-10 (1996).

 
\bibitem{DURHAM} S. Catani, Yu.L. Dokshitzer, M. Olsson, G. Turnock 
        and B.R. Webber, \pl B269 1991 432;\\
%        N. Brown and W.J. Stirling, {\it Phys. Lett.} {\bf B252} (1990) 657;\\
        N. Brown and W.J. Stirling, {\it Z. Phys.} {\bf C53} (1992) 629.

\end{thebibliography}
\end{document}